# Linear gyrokinetic simulation of kinetic infernal mode

Gengxian Li[1], Haotian Chen[2]*, Yong Xiao[1]*

1. Institute for Fusion Theory and Simulation, Zhejiang university, Hangzhou, China
2. Southwestern institute of Physics, Chengdu, China

(Corresponding author) E-mail: chenhaotian@swip.ac.cn
yxiao@zju.edu.cn

## Abstract

Kinetic infernal mode (KIM) is an electromagnetic instability driven by thermal ions in weak magnetic shear region with a frequency similar to the kinetic ballooning mode (KBM). Gyrokinetic simulations of KIM using Gyrokinetic Toroidal Code (GTC) found that the electromagnetic instability shows a smooth transition from KBM to KIM in both frequency and growth rate when magnetic shear varies from strong to weak, which suggests that KIM and KBM may belong to the same mode physically. The mode structure analysis reveals that the mode transition is induced by the change in distance between adjacent mode rational surfaces. The magnetic shear and driving source effects are investigated in detail. The simulation results show that KIM prefers to grow on the mode rational surface nearest to the minimum magnetic shear, i.e., where the shear stabilizing effect is weakest, instead of at the maximum of density gradient or temperature gradient. However, the magnitude of the growth rate is determined by magnetic shear and temperature gradient simultaneously. These findings suggest that KIM can be effectively regulated by modifying the strength and position of magnetic shear, as well as pressure gradients.

## Ⅰ. Introduction

Advanced tokamak scenario, characterized by the high bootstrap current fraction and high confinement, is regarded as one of the most likely choices for modern magnetic confinement fusion devices[1–3]. The possible generation of internal transport barrier (ITB) in this scenario could effectively suppress the ion transport, and then improve plasma confinement over a region of plasma core[4,5]. And this region is often associated with strong $E \times B$ shear flow or a weak magnetic shear $s$ which comes from a flattened safe factor ($q$) profile[6–10]. Previous studies verified that the weak magnetic shear could also suppress the turbulence by generating strong zonal flows[11,12]. Therefore, it is essential to investigate the effect of reversed $q$ profile on fusion plasma instability and associated turbulent transport.

Infernal mode[13,14] is a pressure-driven magnetohydrodynamic (MHD) instability that occurs in the weak or zero magnetic shear region. The normalized plasma pressure $\beta$ for this infernal instability is found below the critical $\beta$ value predicted by the ballooning theory, where $\beta = 8\pi nT/B^2$ is the ratio between plasma thermal energy and magnetic field energy [14]. Extensive research has been conducted on the infernal mode in the past two decades[15–20]. However, most of these studies are within the MHD framework, and there are only few studies of infernal mode including kinetic effects. It is known that the kinetic effects such as finite Larmor radius (FLR) and finite orbit width (FOW) effects have a stabilizing effect on MHD ballooning mode [21,22], which could potentially act as a similar role for the infernal mode. Hence, this instability should be investigated with kinetic effect[23].

It is known that the ballooning representation is not applicable if the instability is in the weak shear region[24]. Thus, theoretical investigation of KIM poses significant challenges due to the inadequacy of conventional ballooning transformation. Meanwhile, the global effect introduced by the sparsity of mode rational surfaces further complicates theoretical analysis of the KIM instability. Therefore, it is crucial to use global gyrokinetic simulation to investigate the physics properties of KIM.

In this paper, we investigate kinetic infernal mode (KIM) with global gyrokinetic simulations by GTC code[25,26]. In practice, KIM can be obtained by changing the q profile from monotonic to reversed magnetic shear in the high β region, while keeping the other parameters the same as those for kinetic ballooning mode (KBM)[27,28]. The growth rate of KIM is found notably larger than that of KBM and their real frequencies are comparable[23]. We simulated the KIM under different q profiles and confirmed here that KIM and KBM may belong to the same mode physically. On the other hand, we conducted an extensive investigation on the relationship between KIM and the stabilizing magnetic shear s, as well as the free energy source, specifically the temperature gradient $(R/L_{ti})$ and density gradient $(R/L_{ni})$, alongside the mode rational surface locations in the simulation domain. The simulation findings show that weak shear is a crucial factor in the formation of KIM and the driving free energy sources only influence the magnitude of its growth rate.

This paper is organized as follows: In Section Ⅱ, the gyrokinetic simulation model in the Gyrokinetic Toroidal Code (GTC) is described. Section Ⅲ presents the mode transition from ion temperature gradient (ITG) to KBM in a normal-shear $q$ profile, and from ITG to KIM in a reversed $q$ profile as $β$ increases. In Section Ⅳ, the relationship between magnetic shear $s$ and KIM is explored through a scan of the magnetic shear $s$. Furthermore, the relationship between KIM and the magnetic shear $s$, as well as the free energy driving sources ($R/L_{ni}$ or $R/L_{Ti}$) are discussed in Section Ⅴ. Finally, in Section Ⅵ, we provide a summary of the KIM simulation results and discuss future work.

## Ⅱ. GTC formulation for KIM simulation

Our simulations are based on the electromagnetic gyrokinetic model for collisionless plasmas implemented in the GTC code. The gyrokinetic equation describes the plasma dynamics in 5D phase space, where gyrocenter position ***X***,

magnetic moment $\mu$, and parallel velocity $v_\parallel$ serve as its coordinates [29–31]:

$$\left(\frac{\partial}{\partial t} + \dot{X} \cdot \nabla + \dot{v}_\parallel \frac{\partial}{\partial v_\parallel}\right) f_\alpha = 0 \tag{1}$$

$$\dot{X} = v_\parallel \frac{\mathbf{B}}{B_0} + v_E + v_c + v_g \tag{2}$$

$$\dot{v}_\parallel = -\frac{1}{m_\alpha} \frac{\mathbf{B}^*}{B_0} \cdot (\mu \nabla B_0 + Z_\alpha \nabla \phi) - \frac{Z_\alpha}{m_\alpha c} \frac{\partial A_\parallel}{\partial t} \tag{3}$$

In the preceding equations, the particle species label $\alpha = i, e$ represents thermal ions and electrons, respectively. In addition, $f_\alpha$ is the distributed function, $m_\alpha$ is the particle mass and $Z_\alpha$ denotes the particle charge. $\mathbf{B} = \mathbf{B_0} + \delta \mathbf{B}$ where $\mathbf{B_0} = B_0 \mathbf{b}$ is the equilibrium magnetic field, and $\delta B = \nabla \times (A_\parallel \mathbf{b})$. $\mathbf{B}^* = \mathbf{B_0^*} + \delta \mathbf{B}$, $\mathbf{B_0^*} = \mathbf{B_0} + \left(\frac{B_0 v_\parallel}{\Omega_\alpha}\right) \nabla \times \mathbf{b}$, with the gyrofrequency $\Omega_\alpha$. $A_\parallel$ and $\phi$ are the gyroaveraged perturbation of electrostatic and vector potentials. $v_E$, $v_c$, $v_g$ is the $E \times B$ drift, curvature drift and grad-B drift, respectively, which are given by

$$\mathbf{v}_E = \frac{c \mathbf{b} \times \nabla \phi}{B_0} \tag{4}$$

$$\mathbf{v}_c = \frac{v_\parallel^2}{\Omega_\alpha} \nabla \times \mathbf{b} \tag{5}$$

$$\mathbf{v}_g = \frac{\mu}{m_\alpha \Omega_\alpha} \mathbf{b} \times \nabla B_0 \tag{6}$$

The continuity equation of adiabatic electron density is given by[26]

$$\frac{\partial \delta n_e}{\partial t} + B_0 \mathbf{b_0} \cdot \nabla \left(\frac{n_{e0} \delta u_{e\parallel}}{B_0}\right) + B_0 v_E \cdot \nabla \left(\frac{n_{e0}}{B_0}\right) - n_{e0}(\mathbf{v_{e*}} + \mathbf{v_E}) \cdot \frac{\nabla B_0}{B_0} \tag{7}$$
$$= 0$$

where $v_{e*} = \mathbf{b} \times \nabla(\delta P_{e\parallel} + \delta P_{e\perp})/(n_{e0} m_e \Omega_e)$, $\delta P_\parallel = \int dv m v_\parallel^2 \delta f_e$, $\delta P_{e\perp} = \int dv \mu B_0 \delta f_e$, $n_{e0} = \int dv f_{e0}$. The electron parallel fluid velocity results from the parallel Ampere's law

$$n_{e0} e \delta u_{\parallel e} = \frac{c}{4\pi} \nabla_\perp^2 \delta A_\parallel + n_{i0} Z_i \delta u_{\alpha \parallel} \tag{8}$$

The vector potential is obtained from the definition of the parallel electric field:

$$\delta E_\parallel = -\mathbf{b_0} \cdot \nabla \phi = \frac{1}{c} \frac{\partial \delta A_\parallel}{\partial t} = -\mathbf{b_0} \cdot \nabla \phi_{eff} \tag{9}$$

$$\frac{\partial \delta A_\parallel}{\partial t} = \nabla_\parallel (\phi_{eff} - \phi) = \nabla_\parallel \phi_{ind} \tag{10}$$

The lowest order $\phi_{eff}$ is calculated by assuming the adiabatic electron response in the ideal MHD limit,

$$\frac{e \phi_{eff}^0}{T_e} = \frac{\delta n_e}{n_{e0}} \tag{11}$$

And $\phi$ is obtained from the gyrokinetic Poisson's equation[32],

$$\frac{Z_i^2 n_i}{T_i}(\phi - \tilde{\phi}) = \sum_{\alpha=i,e} Z_\alpha \delta n_\alpha \qquad (12)$$

The gyrokinetic Vlasov equation together with the parallel Ampere's law, gyrokinetic Poisson's equation, form a closed system for gyrokinetic electromagnetic simulation. We use particle-in-cell (PIC) approach and delta-f method to solve this gyrokinetic-Maxwell plasma system[33]. Defining the ion weight, $w_i = \delta f_i/f_i$, where $f_i = f_{0i} + \delta f_i$, then we have

$$\begin{aligned}\frac{dw_i}{dt} = (1-w_i)[&-\left(v_\parallel \frac{\delta \mathbf{B}}{B} + \mathbf{v_E}\right) \cdot \frac{\nabla f_{0i}}{f_{0i}} \\ &+ \left(\mu \frac{\delta \mathbf{B}}{B_0} \cdot \nabla B_0 + Z_\alpha \frac{\mathbf{B}^*}{B_0} \cdot \nabla \phi + \frac{Z_i}{c} \frac{\partial A_\parallel}{\partial t}\right) \\ &\times \frac{1}{m_\alpha} \frac{1}{f_{0\alpha}} \frac{\partial f_{0\alpha}}{\partial v_\parallel}]\end{aligned} \qquad (13)$$

To consider kinetic electronic effect, the fluid-kinetic hybrid model is adopted in GTC[26]. In this model, the electron response is divided into a lowest order adiabatic part as well as a higher order non-adiabatic part, or a kinetic part. $f_e = f_{0e} + \delta f_e = f_{0e} + \delta f_e^{(0)} + \delta h_e$, the weight is defined as $w_e = \delta h_e/f_e$.

$$\begin{aligned}\frac{dw_e}{dt} = \left(1 - \frac{\delta f_e^{(0)}}{f_{0e}} - w_e\right)[&-\mathbf{v_E} \cdot \nabla ln f_{0e}|_{v_\perp} - \frac{\partial}{\partial t}\frac{\delta f_e^{(0)}}{f_{0e}} - v_d \cdot \nabla \frac{\delta f_e^{(0)}}{f_{0e}} \\ &+ \frac{e}{T_e} v_d \cdot \nabla \phi - \frac{cb_0 \times \nabla \langle \phi \rangle}{B_0} \cdot \nabla \frac{\delta f_e^{(0)}}{f_{0e}} + \frac{ev_\parallel}{cT_e}\frac{\partial \langle A_\parallel \rangle}{\partial t}]\end{aligned} \qquad (14)$$

To manifest the important role of zonal flow, we usually divide the potential $\phi$ and parallel vector potential into to a zonal component and non-zonal component, i.e., $\phi = \delta\phi + \langle\phi\rangle$, $A_\parallel = \delta A_\parallel + \langle A_\parallel \rangle$ [29,34]. However, for present linear instability investigations, we can ignore the contribution from the zonal component and restrict $\phi = \delta\phi$, $A_\parallel = \delta A_\parallel$.

## Ⅲ. Linear instabilities with monotonic and reversed q profiles

In this section, we present linear gyrokinetic simulations for the mode transition in normal and reserved magnetic shear cases with a concentric circular equilibrium specifically as a benchmark. Two distinct profiles are considered: a monotonous normal-shear $q$ profile given by $q(r) = 0.85 + 2.18(r/a)^2$, and a reversed-shear $q$ profile given by $q(r) = 1.76 + 10(r/a - 0.5)^2$, where $a$ represents the minor radius.

The remaining parameters are configured according to the Cyclone Base Case (CBC) parameters: the major radius $R_0 = 83.5cm$, the minor radius $a = 0.36R_0$; the temperature profile is modeled as $T_s = Exp[-T_1 * \tanh((r-r_0)/T_2)]$, with $T_1 = 0.3a/L_t, T_2 = 0.3a$, $r_0 = 0.5a$ representing the reference magnetic surface; similarly, the density profile is expressed as $n = Exp[-n_1 * \tanh((r-r_0)/n_2)]$ with $n_1 = 0.3a/L_n, n_2 = 0.3a$. Such plasma profile settings ensure that $R/L_{ti}(r_0) = 6.92$ and $R/L_n(r_0) = 2.22$. where $L_{ti} = -(dlnT_i/dr)^{-1}$ denotes the temperature gradient scale length and $L_n = -(dlnn/dr)^{-1}$ denotes the temperature gradient scale length, respectively[23,35].

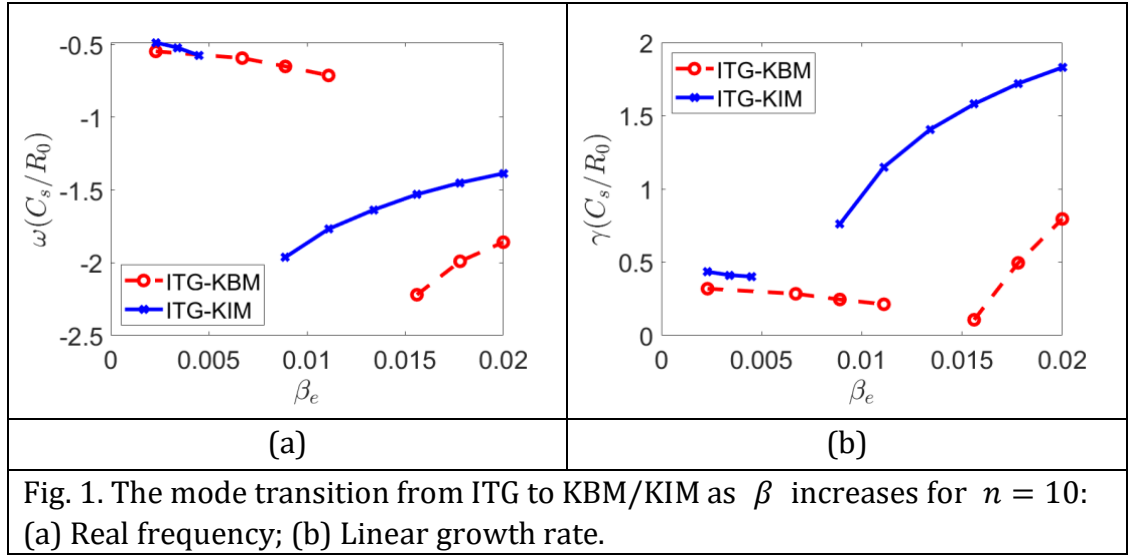

(a)          (b)

Fig. 1. The mode transition from ITG to KBM/KIM as $\beta$ increases for $n = 10$: (a) Real frequency; (b) Linear growth rate.

First, we increase the plasma pressure to observe the mode transition from electrostatic modes to electromagnetic modes for normal and reversed shear cases. Fig.1 displays the linear growth rate and real frequency of this instabilities with $\beta$ for a toroidal mode number $n = 10$. It is evident that as the normalized plasma pressure $\beta$ increases, the real frequency shown in Fig. 1(b) undergoes a jump, indicating the mode has changed from ITG to KIM. For the normal-shear q profile corresponds to the blue line, the mode transition from ITG to KBM occurs at $\beta = 1.4\%$, while in the reversed-shear q profile illustrated by the red line, the threshold value of $\beta$ for mode transition is smaller, occurring at $\beta = 0.8\%$. In the low $\beta$ region, the frequency of ITG remains nearly constant for both cases. However, in the high $\beta$ region ($\beta > 1.2\%$), the growth rate of the KIM is almost twice as large as that of the KBM. Besides, the distinction between KIM and KBM is also evident in mode structures as shown in Fig. 2. KBM has a typically ballooning mode structure, whose different poloidal components are close to the corresponding mode rational surface, respectively. Instead, the mode structure of KIM is local at the weak shear region. With the features of KIM and KBM, the relationship between these electromagnetic modes will be discussed in section Ⅳ.

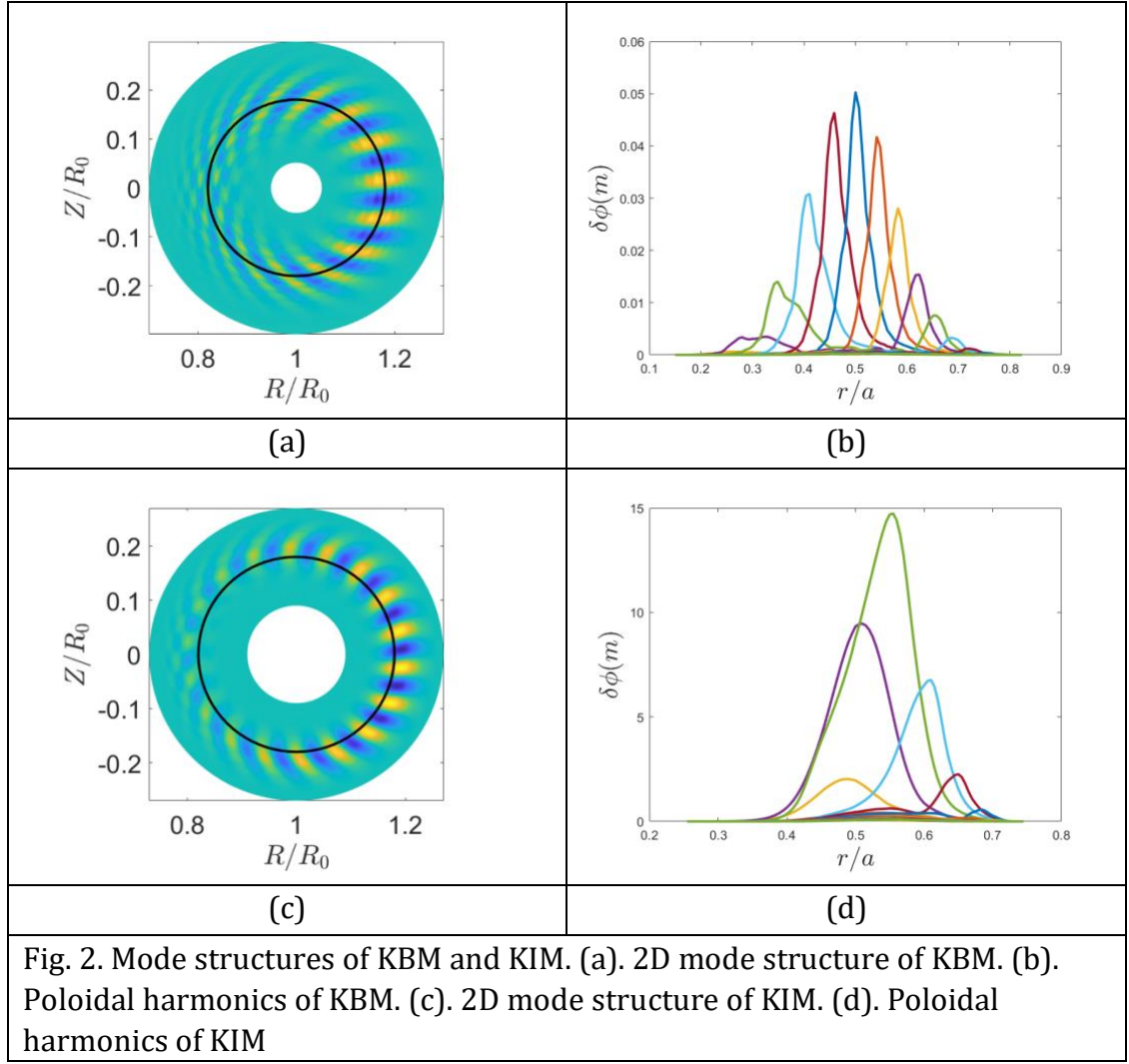

Fig. 2. Mode structures of KBM and KIM. (a). 2D mode structure of KBM. (b). Poloidal harmonics of KBM. (c). 2D mode structure of KIM. (d). Poloidal harmonics of KIM

## Ⅳ. Relationship between KIM and KBM

In last section, we demonstrated that KIM and KBM differ in mode structure, growth rate, and the threshold of $\beta$ for mode transition. In this section, we will explore the connection between KIM and KBM. KIM and KBM are both electromagnetic instabilities driven by pressure gradient, but with different magnetic shears defined as $s = rq'/q$. When $s$ is moderate or strong ($s \sim O(1)$), the pressure gradient driven electromagnetic instability is KBM. However, when s is weak, i.e., $|s| \ll 1$, the pressure gradient driven electromagnetic instability has been discovered to be KIM. Meanwhile, the plasma pressure has to be set sufficiently large to excite electromagnetic instability, thus we set $\beta = 0.02$, and $R/L_{ti} = 6.92$, and $R/L_n = 2.22$. To investigate the relationship between KIM and KBM, we systematically vary the magnetic shear $s$ in the simulation, and measure the frequency and growth rate of the electromagnetic instabilities. In Fig. 3(a), the q profile is represented by $q(r) = 1.76 + c_1 * (r/a - 0.5)^2$, where the slope of q profile or magnetic shear changes as we vary $c_1$ from $-2.0$ to $2.0$. For comparison purposes, it is important to note that the safety factor $q$ at $r = 0.5a$ remains constant at 1.76. Fig. 3(b) illustrates the corresponding magnetic shear

$s(r = 0.5a)$ ranging from $-0.5$ to $0.5$ for different $q$ profiles in Fig. 3(a).

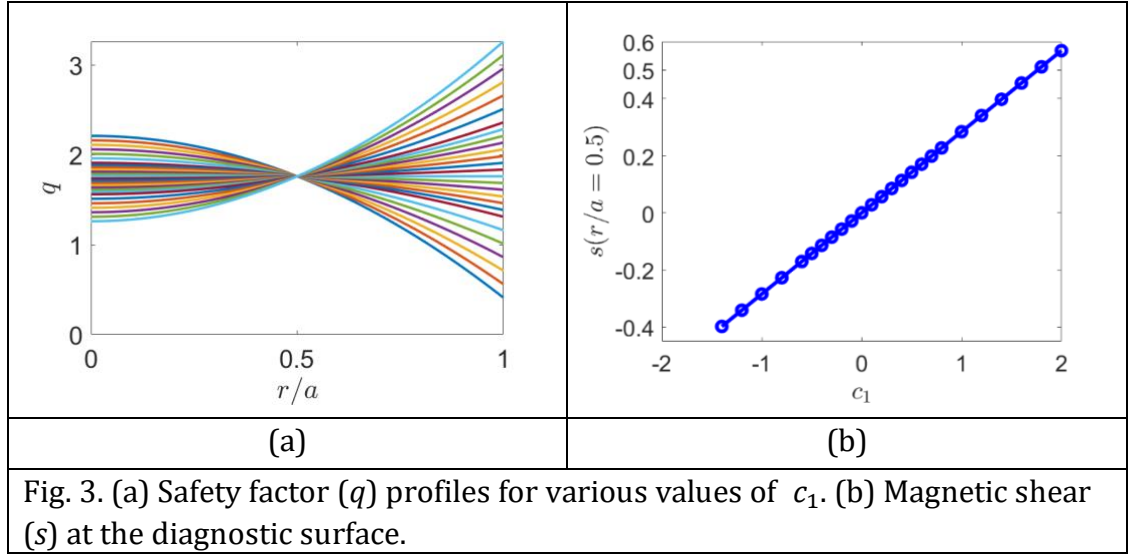

Fig. 3. (a) Safety factor ($q$) profiles for various values of $c_1$. (b) Magnetic shear ($s$) at the diagnostic surface.

Figure. 4 illustrates the frequency with the various q profiles. The growth rate increases initially and then decreases with magnetic shear $s$, with a turning point at approximately $s = 0.2$ in Fig. 4(a). Concurrently, the real frequency demonstrates a non-monotonic trend similar to the growth rate in Fig. 4(b).

To determine whether the instability is KIM or KBM for various magnetic shears, three typical radial mode structures of the perturbative potential $\delta\phi$ are displayed in Fig. 5 for different magnetic shears. The green and blue dashed lines show the scale lengths of the temperature gradient $R/L_{Ti,e}$, and the density gradient $R/L_{ni,e}$, respectively. The red dashed line is the radial profile of the safety factor q. The temperature gradient, density gradient, and q profile are shown in red on the right side of Fig. 4. The solid lines of different colors are the radial mode structures for different poloidal mode numbers (m) with scales shown in the left axis of each subfigure, while the vertical dashed lines of the same color denote the corresponding rational surfaces.

Fig. 5 shows that when the magnetic shear is sufficiently large, i.e., $s \sim 0.4$, the radial profile of different poloidal harmonics for one particular toroidal mode number n exhibits typical characteristics of a ballooning mode structure, suggesting the presence of KBM. However, as the magnetic shear s decreases, the mode rational surfaces become far apart, leading to that the unstable modes no longer grow precisely at the corresponding rational surfaces. This means that the global effect becomes essential for the weak magnetic shears, leading to the breakdown of the ballooning structure[24]. At $s = 0.2$, where the growth rate peaks, the instability is identified as KIM for its mode structure.

Consequently, throughout the parameter scanning of s, there exists a discernible mode transition from KIM to KBM. Unlike the ITG-KBM transition characterized by discontinuous jumps of real frequency, both the real frequency and linear growth rate exhibit continuous changes through the transition from KIM to KBM. Hence, these simulation results indicate that KIM and KBM may belong to the same mode. However, it is worthwhile mentioning that GTC is an initial value code, and thus it is impossible to judge whether the modes belong to the same branch physically in a precise way, and the more definite conclusions should be made by an eigenvalue gyrokinetic code.

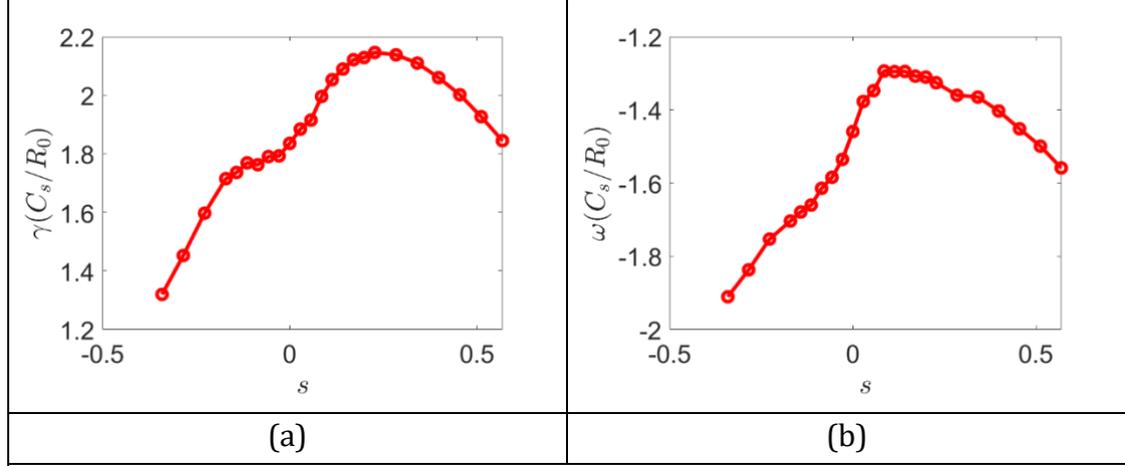

(a)                                (b)

Fig. 4. Linear electromagnetic instabilities for different values of magnetic shear (s) depicted in Fig. 2 with $n = 10$, $\beta = 0.02$: (a) Growth rate; (b) Real frequency.

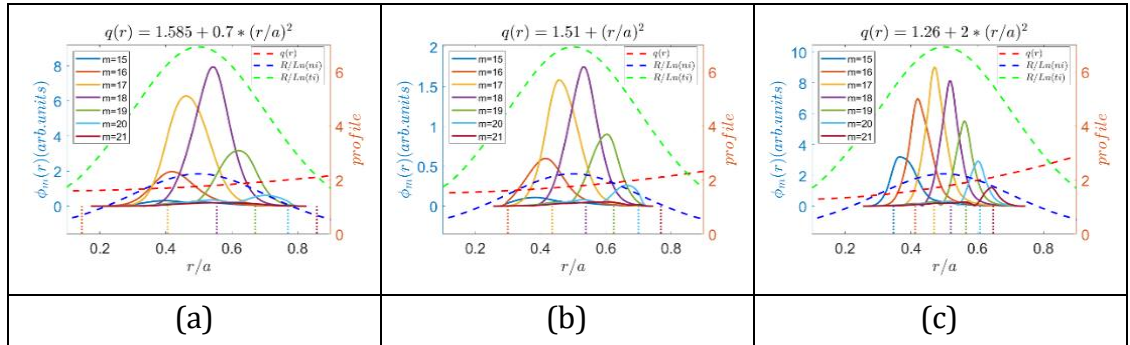

(a)                     (b)                     (c)

Fig. 5. Radial mode structure of electromagnetic instability for different magnetic shears: (a) $s = 0.1989$ (b) $s = 0.2841$ (c) $s = 0.5682$, with $\beta = 0.02$.

To further examine the impact of magnetic shear s on the electromagnetic instabilities, we conduct a simple qualitative analysis of the instabilities with their mode rational surface distribution. We can treat $\delta = \Delta_{\rm rs}/l_e$ as the weak shear condition similar to the strong plasma gradient condition in reference[24], where $l_e$ is the length of the radial mode width, and the $\Delta_{rs} \sim 1/nq'$ represents the distance between adjacent rational surfaces. As depicted in Fig. 5(c), when $s$ is large, $\delta \sim 1/nqs \ll 1$, suggesting that the distance between the mode rational surfaces negligibly small compared to the radial envelope. The use of translational invariance in the ballooning representation is appropriate for treating the coupled poloidal harmonics. In this large $s$ regime, the instability is clearly KBM and exhibits a characteristic ballooning mode structure. As $s$ decreases, an increase in $\delta$ leads to the amplification of the global effect. The modes with different poloidal mode numbers gradually move away from their corresponding rational surfaces, as illustrated in Fig. 5(b). Finally, when $\delta \geq 1$ as observed in Fig.5(a), the variation of radial envelope of the instability is comparable to the variation of a specific poloidal harmonic radial profile with $nq-m$. Therefore, the translational invariance is broken and the unstable mode becomes KIM. In a word, the weakness of magnetic shear leads to the sparsity of mode rational surface, which then bring up the presence of KIM.

## Ⅴ. Physics analysis for kinetic infernal mode

Simulation results presented in the preceding section illustrate that KIM is strongly influenced by the magnetic shear. In general, magnetic shear can influence field line bending term in the vorticity equation, which represents a stabilizing effect according to the energy principle[36], while the instability driving source $d\beta/dr$, which is related to the scale length of temperature and density gradients of ions and electrons, i.e., $R/L_{ni,e}$, $R/L_{ti,e}$ respectively, is represented in the ballooning interchange term in the vorticity equation[21,22,36,37]. In this section, we will examine how these factors, including q profile shape, the magnetic shear and driving source $d\beta/dr$, can affect the linear properties of KIM separately.

### A. Relationship between KIM and safety factor profile

We first examine the relationship between the shape of q profiles and KIM. In fig. 6(a) and fig. 6(d), the reversed q profiles are given as: $q(r) = 1.76 + 2*(r-0.5)^2$ and $q(r) = 1.76 + 20*(r-0.5)^2$ respectively. By comparing 2D mode structure and radial mode structure in Fig6. (b) (c) (e) (f), we can conclude that KIM is clearly affected by the q profile. When the q profile is flat, the mode structure becomes wider, whereas a steep q profile gives a narrower mode structure.

Then we treat q profile a piecewise function in Fig. 7(a) as $q(r) = \begin{cases} 1.76 - 2*(r/a - 0.5)^2 & (r < 0.5a) \\ 1.76 + 30*(r/a - 0.5)^2 & (r > 0.5a) \end{cases}$ and in Fig. 7(d) as $q(r) = \begin{cases} 1.76 - 2*(r/a - 0.5)^2 & (r < 0.5a) \\ 1.76 + 2*(r/a - 0.5)^2 & (r > 0.5a) \end{cases}$. By comparing the two-dimensional mode structure of KIM in fig. 7(b) and fig. 7(e), it appears that the KIM's mode structure has a radial shift. This is due to strong enough stabilizing effect of the magnetic shear $s$ in the region $r > 0.5$ in fig. 7(a), resulting in a small growth rate of KIM.

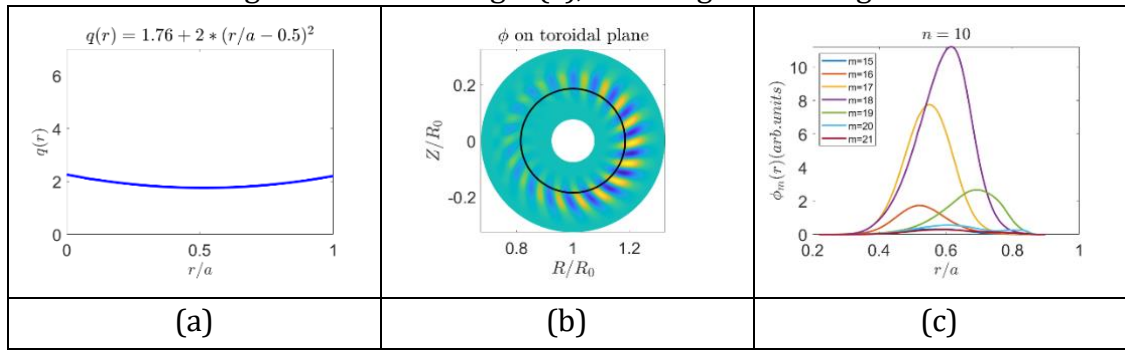

(a)     (b)     (c)

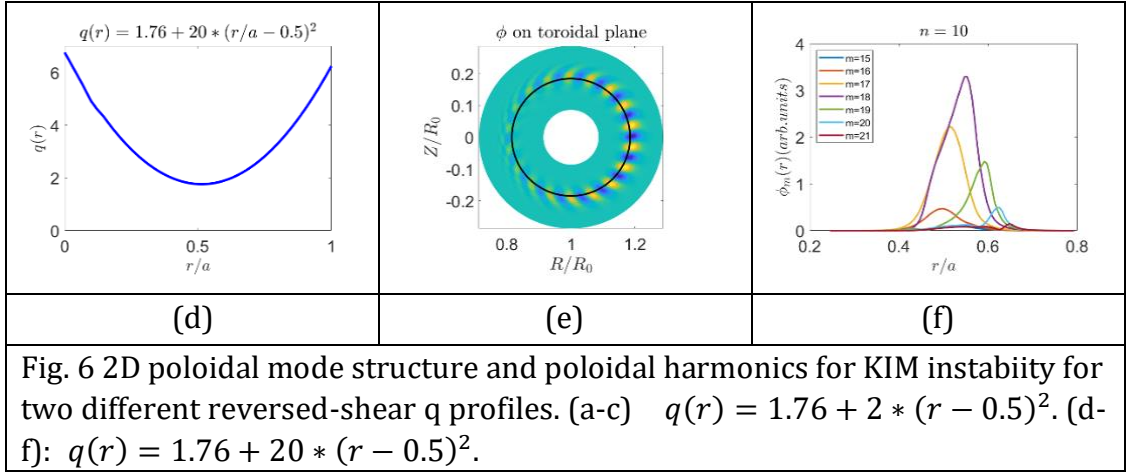

Fig. 6 2D poloidal mode structure and poloidal harmonics for KIM instabiity for two different reversed-shear q profiles. (a-c)  $q(r) = 1.76 + 2*(r - 0.5)^2$. (d-f): $q(r) = 1.76 + 20*(r - 0.5)^2$.

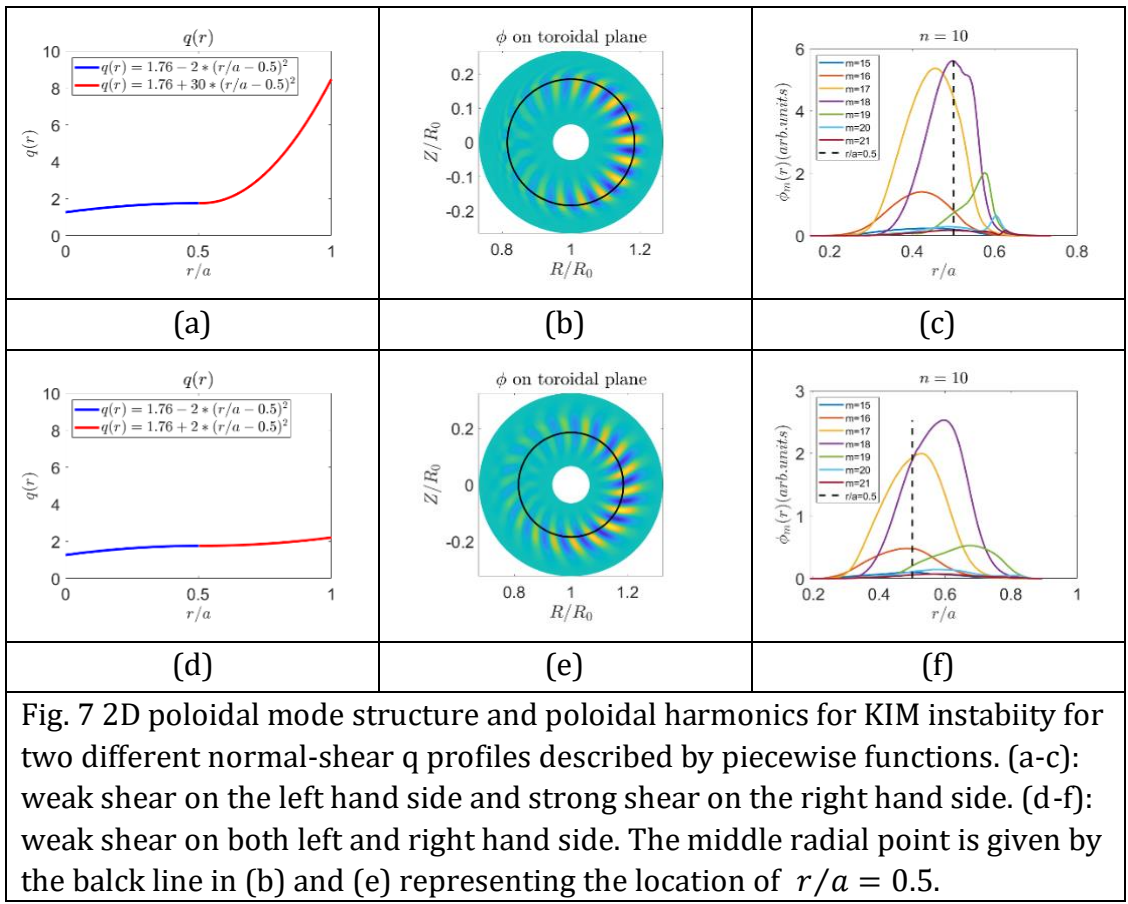

Fig. 7 2D poloidal mode structure and poloidal harmonics for KIM instabiity for two different normal-shear q profiles described by piecewise functions. (a-c): weak shear on the left hand side and strong shear on the right hand side. (d-f): weak shear on both left and right hand side. The middle radial point is given by the balck line in (b) and (e) representing the location of $r/a = 0.5$.

### B. Relationship between KIM and magnetic shear

For the reversed shear configuration, we then investigate how the magnetic shear influences KIM. Fig. (8) presents some cases demonstrating the constraining effect of magnetic shear $s$ on KIM. In Fig. 8, $R/L_{ni,e}$, $R/L_{ti,e}$ are kept constant while the q profile is changed. We set that all q profiles as reversed shear, but the radial positions of $q_{min}$ are shifted from left to right in the simulation domain as shown by the red dash line.

In Fig. 8(a), when the position of $q_{min}$ is close to the left boundary, the q profile appears to be monotonously increasing in the simulation domain, suggesting that the electromagnetic instability is KBM. And in fig .8 (b)-(f), where the q profile is nearly flat, the radial mode structures confirm that the instability is KIM. Due to the weakest stabilizing effect at the location of $q_{min}$, the poloidal harmonics radial profiles of the instabilities move along with the weak shear region. All these simulation results demonstrate that the weak magnetic shear is crucial in formation of KIM.

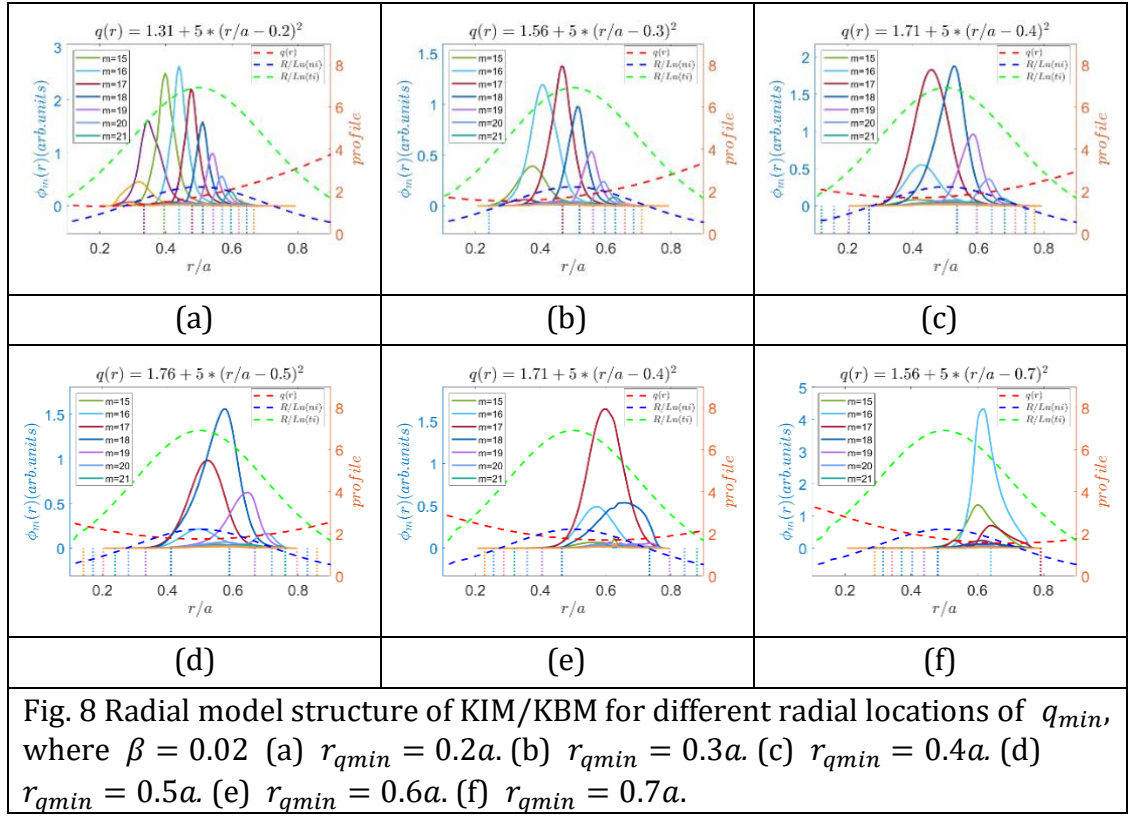

Fig. 8 Radial model structure of KIM/KBM for different radial locations of $q_{min}$, where $\beta = 0.02$ (a) $r_{qmin} = 0.2a$. (b) $r_{qmin} = 0.3a$. (c) $r_{qmin} = 0.4a$. (d) $r_{qmin} = 0.5a$. (e) $r_{qmin} = 0.6a$. (f) $r_{qmin} = 0.7a$.

## C. Relationship between KIM and driving source

Finally, the effects of driving source, especially the scale lengths of temperature and density gradients, on KIM are discussed. Fig. 9 illustrates three cases with different $q$ profiles. In each case, the peaks of $\frac{R}{L_{ti,e}}$ and $\frac{R}{L_n}$ are positioned at the rational surface second closest to the s=0 position, avoiding the influence of the rational surface. Fig. 9(a)-(c) employ CBC parameters, with all simulation results indicating that KIM amplifies at the rational surface closest to the position of $q_{min}$, rather than where the driving source is strongest.

To further study the influence of driving source strength on KIM, stronger driving terms are set as $R/L_{ti,e} = 8.88, 11.1$ with the $q(r) = 1.76 + 2.0 * (r - 0.5)^2$ in Fig. 10(a), (b). The radial peak locations of KIM in Fig. 10(a), (b) align closely with the previous results in Fig. 9. Combining the conclusions of Figs. 8 and 9, it can be inferred that the stabilizing effect from field-line bending, specifically magnetic shear, exerts a stronger influence on the location of KIM compared to the driving source of instability.

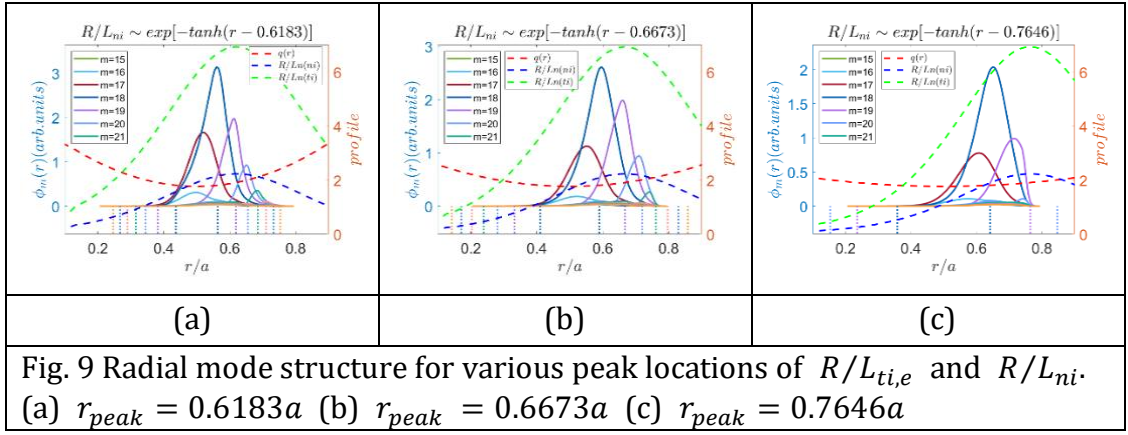

Fig. 9 Radial mode structure for various peak locations of $R/L_{ti,e}$ and $R/L_{ni}$.
(a) $r_{peak} = 0.6183a$ (b) $r_{peak} = 0.6673a$ (c) $r_{peak} = 0.7646a$

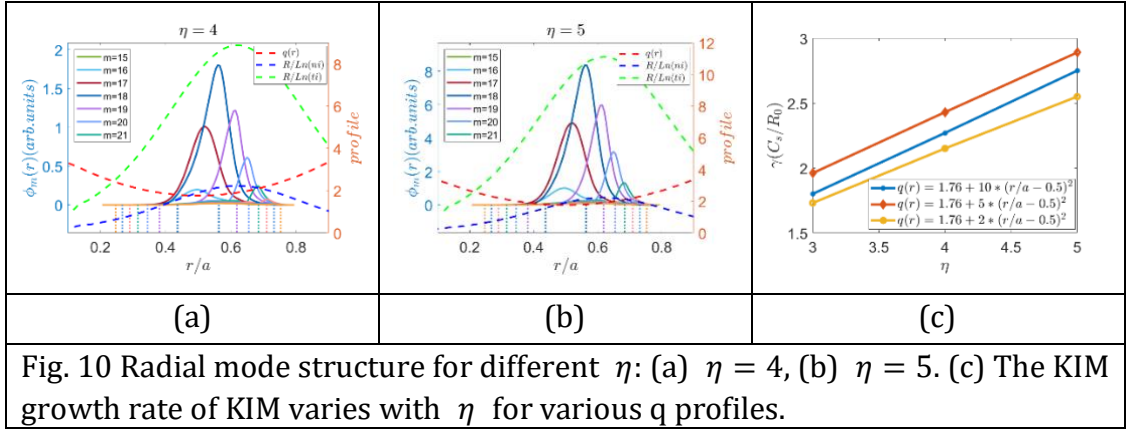

Fig. 10 Radial mode structure for different $\eta$: (a) $\eta = 4$, (b) $\eta = 5$. (c) The KIM growth rate of KIM varies with $\eta$ for various q profiles.

Besides, the KIM growth rate at the diagnostic surface $r = 0.5a$ is assessed with different strength of the driving source for the q profiles illustrated in Fig 8. Fig. 10(c) depicts that, for each $q$ profile, the growth rate of KIM escalates with an increase in the parameter $\eta \equiv L_{ti}/L_{ni}$. This observation suggests that the driving free energy source solely impacts the growth rate of KIM without inducing any changes in its poloidal harmonics (radial mode structures).

## VI. Summary

This study investigates linear properties of KIM using global gyrokinetic simulations with the GTC code. Through varying magnetic shear, the linear growth rate and real frequency varies continuously, suggesting that KIM and KBM may be the same mode physically. A qualitative method has been developed to distinguish KBM and KIM based on the ratio between distance between adjacent mode rational surfaces and radial mode envelope, $\delta$[24]. For KBM, $\delta$<<1, the distance between adjacent mode rational surfaces is negligible compared to the radial envelope, and the global effect is weak due to the strong magnetic shear, resulting in a typical ballooning mode structure. In contrast, when $\delta \geq 1$, KIM locates at the weak shear region, the variation of radial envelope is non-negligible as the distribution of rational surface is sparse, leading to the breaking of translational invariance.

Additionally, the influence of the safety factor profile, the magnetic shear, and the driving source $R/L_{Ti}, R/L_{ni}$ on KIM has been investigated comprehensively.

It is found that the weak shear plays a dominant role in the formation of KIM. And the driven term, pressure gradient primarily affects the growth rate of KIM without altering its radial peak position.

The future work would focus on the nonlinear simulation of KIM to investigate how the zonal flow and zonal current of KIM affect the turbulent transport.

## Acknowledgments

This work was supported by National MCF Energy R & D Program of China under Grant No. 2019YFE03060000, National Natural Science Foundation of China under Grant No. 12375213 & 12375225, and by the Innovation Program of Southwestern Institute of Physics (202301XWCX001).